\documentclass[12pt]{article}

\usepackage{amssymb}
\usepackage{amsmath}
\usepackage{amscd}
\usepackage{mathrsfs}
\usepackage{amssymb}
\usepackage{feynmf}
\usepackage{graphicx}

\def\fslash#1{#1\!\!\!/}

\def\dj{\hbox{d\kern-0.347em \vrule width 0.3em height 1.252ex depth
-1.21ex \kern 0.051em}}

\numberwithin{equation}{section}

\begin{document}

\setlength{\oddsidemargin}{0cm}
\setlength{\baselineskip}{7mm}


\thispagestyle{empty}
\setcounter{page}{0}

\begin{flushright}

\end{flushright}

\vspace*{1cm}

\begin{center}
{\bf \Large Graviton emission in Einstein-Hilbert gravity}



\vspace*{0.3cm}


\vspace*{1cm}

Agust\'{\i}n Sabio Vera$^{a,}$\footnote{\tt 
a.sabio.vera@gmail.com}, 
Eduardo Serna Campillo$^{b,}$\footnote{\tt eduardo.serna@usal.es}, Miguel \' A. V\'azquez-Mozo$^{b,}$\footnote{\tt 
Miguel.Vazquez-Mozo@cern.ch}

\end{center}

\vspace*{0.0cm}

\begin{center}
$^{a}${\sl Instituto de F\'{\i}sica Te\'orica UAM/CSIC \&
Universidad Aut\'onoma de Madrid\\
C/ Nicol\'as Cabrera 15, E-28049 Madrid, Spain
}

$^{b}${\sl Departamento de F\'{\i}sica Fundamental \& IUFFyM,
 Universidad de Salamanca \\ 
 Plaza de la Merced s/n,
 E-37008 Salamanca, Spain
  }
\end{center}

\vspace*{2.5cm}

\centerline{\bf \large Abstract}

\vspace*{0.5cm}

\noindent
The five-point amplitude for the scattering of two distinct scalars with the emission of one graviton in the final state is 
calculated in exact kinematics for Einstein-Hilbert gravity. The result, which satisfies the Steinmann relations, is 
expressed in Sudakov variables, finding that it corresponds to the sum of two gauge invariant contributions written in 
terms of a new two scalar - two graviton effective vertex. A similar calculation is carried out in Quantum Chromodynamics 
(QCD) for the scattering of two distinct quarks with one extra gluon in the final state. The effective vertices which appear in 
both cases are then evaluated in the multi-Regge limit reproducing the well-known result obtained by Lipatov 
where the Einstein-Hilbert graviton emission vertex can be written as the product of two QCD gluon emission vertices, 
up to corrections to preserve the Steinmann relations.

\setcounter{footnote}{0}

\begin{fmffile}{effecvert}

\section{Introduction}

In recent years there has been big progress in the understanding of the structure of scattering amplitudes in 
gauge theories mainly motivated by collider phenomenology but also by the anti de Sitter / conformal field theory (AdS/CFT) 
correspondence~\cite{Maldacena:1997re,Gubser:1998bc,Witten:1998qj}. The latter has boosted the activity towards calculations in the ${\cal N} = 4$ supersymmetric Yang-Mills (MSYM) theory, which 
enjoys four dimensional conformal invariance and allows for calculations up to a large number of quantum loops by 
reducing the problem to a small set of master topologies~\cite{Bern:2010tq}. These results can then be directly used to obtain amplitudes in 
${\cal N} = 8$ supergravity, offering the possibility to investigate the renormalizability of the theory at high orders in the gravitational coupling~\cite{Bern:1998ug,Bern:2009kd}. 

However, there are few results in Einstein-Hilbert gravity~\cite{Donoghue:1994dn,BjerrumBohr:2002kt,Donoghue:1999qh}, where supersymmetry and string theory based 
calculational techniques cannot help~\cite{Dunbar:1994bn,Bern:1993wt,Bern:1991an}. In this case one is forced to approach the calculations with traditional Feynman rules~\cite{Su,Geris:1987ew,Donoghue:1995cz}. 
At each order of perturbation theory the task at hand grows tremendously due to the new multi-graviton vertices appearing 
and the problem becomes a combinatorial nightmare. To make progress in this direction it is needed to find simplifying 
techniques besides using computer algebra. 

It is also possible to get a glimpse of the all orders amplitudes if they are evaluated 
in certain limits, with a remarkable example being the work of Lipatov evaluating multi-graviton scattering processes in 
multi-Regge kinematics (MRK). In this region the amplitudes can be written in a factorized form which allows even for the 
construction of a high energy effective action from which to generate them~\cite{Lipatov:2011ab}. 
The key ingredients in these calculations are the 
reggeization of the graviton~\cite{Grisaru:1975tb,Grisaru:1981ra}  
 together with a full control of eikonal and double logarithmic 
 contributions~\cite{Lipatov:1982vv,Lipatov:1982it,Lipatov:1991nf}. A remarkable result in  Lipatov's investigations is that the graviton emission effective vertex can be written as a double copy of the gluon emission 
effective vertex when both are evaluated in MRK and the latter is calculated in 
QCD~\cite{BFKL1,BFKL2,BFKL3}. 

In the present work we take a bottom-up approach and evaluate inelastic amplitudes at tree level both in Einstein-Hilbert gravity 
and QCD. We obtain expressions 
valid in general kinematics and then go to MRK to reproduce Lipatov's results both in gravity and QCD. 
Our computations are performed using conventional Feynman rules, with the aid of computer 
algebra~\cite{xperm} in 
the much more complicated case of gravity. 

In Section 2 a five-point 
amplitude is evaluated in QCD with two pairs of distinct quarks plus an emitted graviton. We split the contribution of the 
three gluon vertex into two pieces which, when added to the diagrams corresponding to gluon emission from the same 
fermion line, generate an effective vertex which is gauge invariant. In this way the amplitude can be simply decomposed into the sum 
of two topologies constructed with this effective vertex. The MRK limit of this sum coincides with Lipatov's gluon emission vertex. In Section 3 a similar calculation is performed in the case of Einstein-Hilbert gravity, where now the 
five-point amplitude consists of two pairs of distinguishable scalars and one emitted graviton. Operating in 
de Donder gauge we write the result for the exact amplitude in a Sudakov decomposition of the participating 
momenta. Similarly to the QCD case, a new effective vertex appears which allows to write the amplitude as 
the sum of two gauge invariant topologies written in terms of it. This simplification occurs only after noticing 
a novel and subtle partial cancelation between the two diagrams containing the two scalar - two graviton 
vertex, and splitting the three graviton vertex into two different pieces. When taking the MRK limit of the 
sum of these two new effective diagrams we recover Lipatov's results for the graviton emission 
vertex~\cite{Lipatov:1982vv,Lipatov:1982it,Lipatov:1991nf}.  
We also take the MRK limit of our exact calculation to check that 
it can indeed be written as the square of the QCD gluon emission vertex previously 
calculated, plus a contribution needed to keep consistency with the Steinmann relations. This structure is deeply connected to the proposal that gravity can be understood as a ``double copy" of a gauge theory (see {\it e.g.}~\cite{Bern:2002kj} for a review on the subject). Finally, in the 
Conclusions we summarize our main results and provide suggestions for future research.

\section{Quark-quark scattering with gluon emission}

We start the presentation of our work by describing in this section the scattering at tree level 
of two distinct quarks with the emission of a gluon in the final state. Our notation for the different momenta 
reads
\begin{eqnarray}
Q(p,j)+Q'(q,n)\longrightarrow Q(p',i)+Q'(q',m)+g(k,a),
\end{eqnarray}
where $j,n,i,m$ denote the (fundamental representation)
gauge indices of the incoming and outgoing quarks and $a$ the index of the outgoing gluon.
At leading order in the strong coupling constant, $g$, five diagrams contribute to the total amplitude:
\begin{eqnarray}
\nonumber \\[0.2cm]
{M}&\equiv&\hspace*{1cm}
\parbox{42mm}{
\begin{fmfgraph*}(90,60)
\fmfbottom{i1,d1,o1}
\fmftop{i2,d2,o2}
\fmfright{r1}
\fmflabel{$(p,j)$}{i2}
\fmflabel{$(p',i)$}{o2}
\fmflabel{$(k,a)$}{r1}
\fmflabel{$(q,n)$}{i1}
\fmflabel{$(q',m)$}{o1}
\fmf{fermion}{i1,v1}
\fmf{plain}{v1,v2}
\fmf{fermion}{v2,o1}
\fmf{fermion}{i2,v3,v4,o2}
\fmf{phantom}{i1,o1}
\fmf{gluon,tension=0}{v1,v3}
\fmf{gluon,tension=0}{v4,r1}
\end{fmfgraph*} 
}\hspace*{0.5cm}+\hspace*{1.3cm}
\parbox{42mm}{
\begin{fmfgraph*}(90,60)
\fmfbottom{i1,d1,o1}
\fmftop{i2,d2,o2}
\fmfright{r1}
\fmflabel{$(p,j)$}{i2}
\fmflabel{$(p',i)$}{o2}
\fmflabel{$(k,a)$}{r1}
\fmflabel{$(q,n)$}{i1}
\fmflabel{$(q',m)$}{o1}
\fmf{fermion}{i1,v1}
\fmf{plain}{v1,v2}
\fmf{fermion}{v2,o1}
\fmf{phantom}{i1,o1}
\fmf{fermion}{i2,v3,v4,o2}
\fmf{gluon,tension=0}{v2,v4}
\fmf{gluon,tension=0,rubout}{v3,r1}
\end{fmfgraph*} 
} \nonumber \\[1.2cm]
& & \hspace*{-0.4cm} +\hspace*{1cm}
\parbox{42mm}{
\begin{fmfgraph*}(90,60)
\fmfbottom{i1,d1,o1}
\fmftop{i2,d2,o2}
\fmfright{r1}
\fmflabel{$(p,j)$}{i2}
\fmflabel{$(p',i)$}{o2}
\fmflabel{$(k,a)$}{r1}
\fmflabel{$(q,n)$}{i1}
\fmflabel{$(q',m)$}{o1}
\fmf{fermion}{i1,v1,v2,o1}
\fmf{phantom}{i1,o1}
\fmf{fermion}{i2,v3}
\fmf{plain}{v3,v4}
\fmf{fermion}{v4,o2}
\fmf{gluon,tension=0}{v1,v3}
\fmf{gluon,tension=0,rubout}{r1,v2}
\end{fmfgraph*} 
}\hspace*{0.5cm}+\hspace*{1.3cm}
\parbox{42mm}{
\begin{fmfgraph*}(90,60)
\fmfbottom{i1,d1,o1}
\fmftop{i2,d2,o2}
\fmfright{r1}
\fmflabel{$(p,j)$}{i2}
\fmflabel{$(p',i)$}{o2}
\fmflabel{$(k,a)$}{r1}
\fmflabel{$(q,n)$}{i1}
\fmflabel{$(q',m)$}{o1}
\fmf{fermion}{i1,v1,v2,o1}
\fmf{phantom}{i1,o1}
\fmf{fermion}{i2,v3}
\fmf{plain}{v3,v4}
\fmf{fermion}{v4,o2}
\fmf{gluon,tension=0}{v2,v4}
\fmf{gluon,tension=0,rubout}{r1,v1}
\end{fmfgraph*} 
}
\label{eq:feyn_diag_qcd}
\nonumber\\[1.2cm]
& & 
\hspace*{2.5cm}
+\hspace*{1cm}
\parbox{42mm}{
\begin{fmfgraph*}(90,60)
\fmfbottom{i1,d1,o1}
\fmftop{i2,d2,o2}
\fmfright{r1}
\fmflabel{$(p,j)$}{i2}
\fmflabel{$(p',i)$}{o2}
\fmflabel{$(q,n)$}{i1}
\fmflabel{$(q',m)$}{o1}
\fmf{fermion}{i1,v1,o1}
\fmf{fermion}{i2,v3,o2}
\fmf{gluon,tension=0}{v1,v3}
\fmf{phantom}{i1,o1}
\end{fmfgraph*} 
}\hspace*{-2.4cm}
\parbox{42mm}{
\begin{fmfgraph*}(40,20)
\fmfleft{i1}
\fmfright{o1}
\fmflabel{$(k,a)$}{o1}
\fmf{gluon,tension=1}{i1,o1}
\end{fmfgraph*} 
}
\hspace*{-2cm} 
\\
\nonumber
\end{eqnarray}
whose respective contributions we denote by
\begin{eqnarray}
M=M_{1}+M_{2}+M_{3}+M_{4}+M_{5}.
\end{eqnarray}
The evaluation of the individual diagrams gives the result \cite{Gastmans:1990xh}
\begin{eqnarray}
M_{1}&=& -{g^{3}\over 2t'}\overline{u}(p')\fslash{\varepsilon} T^{a}_{ik}{\fslash{p}{}'+\fslash{k}\over 
p'\cdot k}\gamma^{\mu}T^{b}_{kj}u(p)\overline{u}(q')\gamma_{\mu}T^{b}_{mn}u(q), \\
M_{2}&=& {g^{3}\over 2t'}\overline{u}(p')\gamma^{\mu}T^{b}_{ik}{\fslash{p}-\fslash{k}\over p\cdot k}
\fslash{\varepsilon}T^{a}_{kj}u(p)\overline{u}(q')\gamma_{\mu}T^{b}_{mn}u(q), \\
M_{3}&=& -{g^{3}\over 2t}\overline{u}(p')\gamma^{\mu}T^{b}_{ij}u(p)\overline{u}(q')\fslash{\varepsilon}T^{a}_{mk}
{\fslash{q}'+\fslash{k}\over q'\cdot k}\gamma_{\mu}T^{b}_{kn}u(q), \\
M_{4} &=& {g^{3}\over 2t}\overline{u}(p')\gamma^{\mu}T^{b}_{ij}u(p)\overline{u}(q')\gamma_{\mu}
T^{b}_{mk}{\fslash{q}-\fslash{k}\over q\cdot k}\fslash{\varepsilon}T^{a}_{kn}u(q),  \\
M_{5}&=& -{ig^{3}\over tt'}\overline{u}(p')\gamma^{\mu}T^{c}_{ij}u(p)\overline{u}(q')\gamma^{\nu}T^{b}_{mn}u(q)
f^{cba}\nonumber \\
&\times&\Big[(p-p'-q+q')\cdot\varepsilon\,\eta_{\mu\nu} +(q-q'+k)_{\mu}\varepsilon_{\nu}+(p'-k-p)_{\nu}\varepsilon_{\mu}\Big], 
\label{eq:all_contributions}
\end{eqnarray}
where 
\begin{eqnarray}
t=(p-p')^{2},\hspace*{1cm} t'=(q-q')^{2}.
\end{eqnarray}
Following \cite{Xu:1986xb}, we decompose the amplitude into two sets
\begin{eqnarray}
M_{\uparrow}=M_{1}+M_{2}+M_{5}', \hspace*{1cm} 
M_{\downarrow}= M_{3}+M_{4}+M_{5}'',
\end{eqnarray}
where
\begin{eqnarray}
M_{5}'={t\over t-t'}M_{5}, \hspace*{1cm} M_{5}''={t'\over t'-t}M_{5}.
\end{eqnarray}
The full amplitude is then written as
\begin{eqnarray}
M=M_{\uparrow}+M_{\downarrow}.
\label{eq:m-decomposition}
\end{eqnarray}

What makes this decomposition interesting is 
that $M_{\uparrow}$ and $M_{\downarrow}$ are gauge invariant separately. Indeed, replacing the
external polarization $\epsilon_{\mu}(k)$ by $k_{\mu}$ we find that, after some algebra,
\begin{eqnarray}
M_{1}+M_{2}\longrightarrow -{ig^{3}\over  t'}f^{abc}T^{c}_{ij}T^{b}_{mn}
\overline{u}(p')\gamma^{\mu}u(p)
\overline{u}(q')\gamma_{\mu}u(q),
\label{eq:gauge_m1m2}
\end{eqnarray}
whereas 
\begin{eqnarray}
M_{5}'\longrightarrow 
{ig^{3}\over t'(t-t')}k\cdot (p-p'-q+q')
f^{abc}T^{c}_{ij}T^{b}_{mn}\overline{u}(p')\gamma^{\mu}u(p)\overline{u}(q')\gamma_{\mu}u(q).
\end{eqnarray}
When momentum conservation is imposed this last term cancels Eq.~\eqref{eq:gauge_m1m2} and leads to the partial Ward identity
\begin{eqnarray}
k_{\mu}M_{\uparrow}^{\mu}=0, 
\end{eqnarray}
where the notation $M_{\uparrow}\equiv\varepsilon_{\mu}(k)M_{\uparrow}^{\mu}$ has been used.
A similar treatment of the second combination of diagrams generates the Ward identity
\begin{eqnarray}
k_{\mu}M_{\downarrow}^{\mu}=0.
\end{eqnarray}

The main advantage of combining Feynman diagrams into these gauge invariant contributions is that it allows 
to fix the polarization of the emitted gluons independently for each of them, reducing the number of cross terms when evaluating higher order amplitudes by means of unitarity relations.  In the present work the 
existence of these two gauge invariant combinations offers the possibility to define effective vertices which 
can, eventually, be obtained from a suitable effective action valid for general kinematics. In order to extract these vertices from our 
representation of the amplitude we write the two subamplitudes as
\begin{eqnarray}
M_{\uparrow}&=&
\Big[\varepsilon_{\nu}(k)\overline{u}(p')\Lambda^{\mu,\nu}_{ij;c,a}(p,p',k)u(p)\Big]\,{-i\eta_{\mu\sigma}
\delta_{cb}\over t'}
\Big[\overline{u}(q')
g\gamma^{\sigma}T^{b}_{mn}u(q)\Big], 
\label{effective_vertex_form_up}\\
M_{\downarrow}&=&
\Big[\overline{u}(p')g\gamma^{\mu}T^{c}_{ij}u(p)\Big]{-i\eta_{\mu\sigma}\delta_{cb}\over t}
\Big[\varepsilon_{\nu}(k)\overline{u}(q'){\Lambda}^{\sigma,\nu}_{ij;b,a}(q,q',k)u(q)\Big],
\label{effective_vertex_form_down}
\end{eqnarray}
where $\Lambda^{\mu,\nu}_{ij;a,b}(p,p',k)$ is given by
\begin{eqnarray}
\Lambda^{\mu,\nu}_{ij;d,a}&=&
-{ig^{2}\over 2}T^{a}_{ik}T^{d}_{kj}\gamma^{\nu} {\fslash{p}{}'+\fslash{k}\over 
p'\cdot k}\gamma^{\mu}+
{ig^{2}\over 2}T^{d}_{ik}T^{a}_{kj}\gamma^{\mu}{\fslash{p}-\fslash{k}\over p\cdot k}
\gamma^{\nu} \nonumber\\
&&\hspace{-2cm}-{g^{2}\over 2k\cdot (p-p')}f^{adc}T^{c}_{ij}\gamma^{\alpha}
\Big[(2p-2p'-k)^{\nu}\,\delta_{\alpha}^{\mu}
+(p'-p+2k)_{\alpha}\eta^{\mu\nu}+(p'-p-k)^{\mu}\delta^{\nu}_{\alpha}\Big].
\end{eqnarray}
By construction it satisfies the gauge Ward identity
\begin{eqnarray}
k_{\nu}\Lambda^{\mu,\nu}_{ij;a,b}(p,p',k)=0. 
\end{eqnarray}

In writing this effective vertex a notation stressing the nonequivalence of the indices has been used.
Its non-locality is manifest by the presence of momenta in the denominators. 
Diagrammatically, we represent it by
\begin{eqnarray}
\nonumber \\[-0.2cm]
\label{effverup}
{\Lambda}^{\mu,\nu}_{ij;a,b}(p,p',k)=\hspace*{1cm}
\parbox{42mm}{
\begin{fmfgraph*}(90,60)
\fmfblob{.15w}{v3}
\fmfbottom{i1,d1,o1}
\fmftop{i2,d2,o2}
\fmfright{r1}
\fmflabel{$p$}{i2}
\fmflabel{$a,\nu,k$}{r1}
\fmflabel{$p'$}{o2}
\fmflabel{$b,\mu,p'-p-k$}{v1}
\fmf{phantom}{i1,v1,o1}
\fmf{fermion}{i2,v3,o2}
\fmf{gluon,tension=0}{v3,v1}
\fmf{gluon,tension=0}{v3,r1}
\end{fmfgraph*} 
} \\[0.2cm]
\nonumber
\end{eqnarray}
where it should be kept in mind that, by definition, three of the four legs in the vertex 
(those labelled by $p$, $p'$ and $k$) are on-shell, $p^{2}=p'{}^{2}=k^{2}=0$. 
The process $QQ'\rightarrow QQ'g$ at tree level can thus be represented in terms of just two Feynman diagrams, each containing a single effective vertex, {\it i.e.}
\begin{eqnarray}
\nonumber \\[-0.2cm]
M=\hspace*{1cm}
\parbox{42mm}{
\begin{fmfgraph*}(90,60)
\fmfblob{.15w}{v3}
\fmfbottom{i1,d1,o1}
\fmftop{i2,d2,o2}
\fmfright{r1}
\fmflabel{$p$}{i2}
\fmflabel{$q$}{i1}
\fmflabel{$k$}{r1}
\fmflabel{$p'$}{o2}
\fmflabel{$q'$}{o1}
\fmf{fermion}{i1,v1,o1}
\fmf{fermion}{i2,v3,o2}
\fmf{gluon,tension=0}{v3,v1}
\fmf{gluon,tension=0}{v3,r1}
\end{fmfgraph*} 
}+\hspace*{1cm}
\parbox{42mm}{
\begin{fmfgraph*}(90,60)
\fmfblob{.15w}{v1}
\fmfbottom{i1,d1,o1}
\fmftop{i2,d2,o2}
\fmfright{r1}
\fmfright{r1}
\fmflabel{$p$}{i2}
\fmflabel{$q$}{i1}
\fmflabel{$k$}{r1}
\fmflabel{$p'$}{o2}
\fmflabel{$q'$}{o1}
\fmf{fermion}{i1,v1,o1}
\fmf{fermion}{i2,v3,o2}
\fmf{gluon,tension=0}{v3,v1}
\fmf{gluon,tension=0}{v1,r1}
\end{fmfgraph*} 
}
\label{eq:full_diagram_effecver} \\[0.1cm] \nonumber
\end{eqnarray}
This decomposition of the amplitude in terms of two topologies with a single non-local effective vertex is exact, {\it i.e.} independent of the particular
kinematical regime considered. 

\subsection{Multi-Regge kinematics}

The expression for our effective vertex simplifies when restricting the amplitudes to 
multi-Regge kinematics (MRK), where the limit $s=(p+q)^{2} \rightarrow \infty$ is taken while the 
momentum transfers $t$ and $t'$ are kept fixed, {\it i.e.} not growing with energy. These conditions translate 
into the constraints
\begin{eqnarray}
s\gg s_{p'k},s_{q'k}\gg t  \sim t',
\end{eqnarray}
where $s_{p'k}=(p'+k)^{2}$ and $s_{q'k}=(q'+k)^{2}$. This generalizes the 
standard Regge limit in four-point amplitudes to the case of five-point amplitudes.

To implement MRK it is convenient to introduce the momenta $k_{1}$ and $k_{2}$ to write 
\begin{eqnarray}
p'=p-k_{1}, \hspace*{1cm} q'=q+k_{2}, \hspace*{1cm} k=k_{1}-k_{2}
\label{eq:new_momenta}
\end{eqnarray}
and express them using the Sudakov parametrization of the form 
\begin{eqnarray}
k_{1}^{\mu} = \alpha_{1}p^{\mu}+\beta_{1}q^{\mu}+k_{1,\perp}^{\mu}, \hspace{1cm}
k_{2}^{\mu} = \alpha_{2}p^{\mu}+\beta_{2}q^{\mu}+k_{2,\perp}^{\mu},
\label{eq:sudakov_paramet}
\end{eqnarray}
where $p^2=q^2=p \cdot k_{i,\perp} = 0$. 
In terms of the Sudakov parameters $\alpha_{1,2}$ and $\beta_{1,2}$ the multi-Regge limit reads
\begin{eqnarray}
1 \gg \alpha_1 \gg \alpha_2 = \frac{-t'}{s}, \hspace*{1cm} 1 \gg |\beta_2| \gg |\beta_1| = \frac{-t}{s},
\label{MRK1}
\end{eqnarray}
which, for the emitted gluon, implies that
\begin{eqnarray}
k^\mu \simeq \frac{s_{q' k}}{s} p_1^\mu + \frac{s_{p' k}}{s} p_2^\mu + k_\perp^\mu, \hspace{1cm}
k^2 \simeq \frac{s_{p' k} s_{q' k}}{s} + k_\perp^2 = 0.
\end{eqnarray}
The MRK limit, taken on the different contributions to the total amplitude, 
gives the following expression for the effective vertex in Eq.~\eqref{effverup}:
\begin{eqnarray}
\nonumber \\[-0.2cm]
\left.\left(\hspace*{0.5cm}\parbox{42mm}{
\begin{fmfgraph*}(90,60)
\fmfblob{.15w}{v3}
\fmfbottom{i1,d1,o1}
\fmftop{i2,d2,o2}
\fmfright{r1}
\fmflabel{$p$}{i2}
\fmflabel{$a,\nu,k_{1}-k_{2}$}{r1}
\fmflabel{$p-k_{1}$}{o2}
\fmflabel{$b,\mu,k_{2}$}{v1}
\fmf{phantom}{i1,v1,o1}
\fmf{fermion}{i2,v3,o2}
\fmf{gluon,tension=0}{v3,v1}
\fmf{gluon,tension=0}{v3,r1}
\end{fmfgraph*} 
} \hspace*{1.3cm}\right)\right|_{\rm MRK}\hspace*{7cm}
\nonumber \\ 
\nonumber
\\[0.4cm]
= {2g^{2}\over t-t'}f^{abc}T^{c}_{ij}p^{\mu}
\left[\left(
\alpha_{1}-2{t-t'\over s\beta_{2}}\right)p^{\nu}+\beta_{2}q^{\nu}
-(k_{1}+k_{2})^{\nu}_{\perp}\right].
\label{eq:upper_regge_vertex_gauge}
\end{eqnarray}
According to Eqs. \eqref{effective_vertex_form_up} and \eqref{effective_vertex_form_down},  
to use the effective vertex in the second diagram in 
\eqref{eq:full_diagram_effecver} one needs to replace 
$p^{\mu}\rightarrow q^{\mu}$ and $p'{}^{\mu}\rightarrow q'{}^{\mu}$, 
with $k^{\mu}$ unchanged. In terms of the momenta and Sudakov variables 
appearing in the MRK vertex \eqref{eq:upper_regge_vertex_gauge}, this amounts to
\begin{eqnarray}
p^{\mu} \longleftrightarrow q^{\mu}, \hspace*{1cm} 
\alpha_{1} \longleftrightarrow -\beta_{2}, \hspace*{1cm}  
k_{1\perp}^{\mu}\longleftrightarrow -k_{2\perp}^{\mu}.
\end{eqnarray}

To recover Lipatov's reggeized gluon - reggeized gluon - gluon effective vertex at leading order in MRK we 
simply add the contributions of $M_{\uparrow}$ and $M_{\downarrow}$ in this limit. Diagrammatically
\begin{eqnarray}
\nonumber \\[-0.3cm]
\left.\left(\hspace*{1cm}\parbox{35mm}{
\begin{fmfgraph*}(80,50)
\fmfblob{.15w}{v3}
\fmfbottom{i1,d1,o1}
\fmftop{i2,d2,o2}
\fmfright{r1}
\fmflabel{$p$}{i2}
\fmflabel{$q$}{i1}
\fmflabel{$k$}{r1}
\fmflabel{$p'$}{o2}
\fmflabel{$q'$}{o1}
\fmf{fermion}{i1,v1,o1}
\fmf{fermion}{i2,v3,o2}
\fmf{gluon,tension=0}{v3,v1}
\fmf{gluon,tension=0}{v3,r1}
\end{fmfgraph*} 
}+\hspace*{0.5cm}
\parbox{38mm}{
\begin{fmfgraph*}(80,50)
\fmfblob{.15w}{v1}
\fmfbottom{i1,d1,o1}
\fmftop{i2,d2,o2}
\fmfright{r1}
\fmfright{r1}
\fmflabel{$p$}{i2}
\fmflabel{$q$}{i1}
\fmflabel{$k$}{r1}
\fmflabel{$p'$}{o2}
\fmflabel{$q'$}{o1}
\fmf{fermion}{i1,v1,o1}
\fmf{fermion}{i2,v3,o2}
\fmf{gluon,tension=0}{v3,v1}
\fmf{gluon,tension=0}{v1,r1}
\end{fmfgraph*} 
}\right)\right|_{\rm MRK}
=
\hspace*{0.5cm}
\parbox{30mm}{
\begin{fmfgraph*}(80,50)
\fmfbottom{i1,d1,o1}
\fmftop{i2,d2,o2}
\fmfright{r1}
\fmflabel{$p$}{i2}
\fmflabel{$p'$}{o2}
\fmflabel{$q$}{i1}
\fmflabel{$q'$}{o1}
\fmf{fermion}{i1,v1,o1}
\fmf{fermion}{i2,v3,o2}
\fmf{phantom,tension=0}{v1,v3}
\fmf{phantom}{i1,o1}
\end{fmfgraph*} 
}\hspace*{-2.9cm}
\parbox{30mm}{
\begin{fmfgraph*}(70,40)
\fmfv{decor.shape=circle,decor.filled=full,decor.size=.15w}{v1}
\fmfbottom{b1}
\fmfleft{l1}
\fmftop{t1}
\fmfright{r1}
\fmflabel{$k$}{r1}
\fmf{gluon}{b1,v1,t1}
\fmf{gluon}{v1,r1}
\fmf{phantom}{l1,v1}
\end{fmfgraph*} 
} ,
\\[-0.1cm] \nonumber
\end{eqnarray}
where the last effective Feynman diagram is given by 
\begin{eqnarray}
M= \varepsilon_{\nu}(k_{1}-k_{2})\Big(2gp^{\mu}T^{c}_{ij}\Big)
\left({-i\over t}\right)f^{cab}\Gamma^{\nu}_{\mu\sigma}
\left({-i\over t'}\right)\Big(2gq^{\sigma}T^{b}_{mn}\Big),
\end{eqnarray}
with
\begin{eqnarray}
\Gamma^{\nu}_{\mu\sigma}=ig\eta_{\mu\sigma}\left[\left(\alpha_{1}-{2t\over s\beta_{2}}\right)p^{\nu}
+\left(\beta_{2}-{2t'\over s\alpha_{1}}\right)q^{\nu}-(k_{1}+k_{2})_{\perp}^{\nu}\right].
\label{QCDlevvertex}
\end{eqnarray}
This vertex is well-known and, when used to construct elastic amplitudes together with the gluon 
Regge trajectory, it generates the evolution Hamiltonian of the Balitsky-Fadin-Kuraev-Lipatov 
equation~\cite{BFKL1,BFKL2,BFKL3}. This Hamiltonian presents holomorphic 
separability~\cite{Lev1} and invariance under SL(2,$\mathbb{C}$) transformations~\cite{Lipatov:1985uk}. Its 
generalization to the exchange of an arbitrary number of reggeized gluons in the 
$t$-channel (BKP equation~\cite{Bartels:1980pe,Kwiecinski:1980wb}) can be mapped into an 
integrable and symmetric under duality~\cite{Lipatov:1998as} periodic XXX Heisenberg 
ferromagnet~\cite{Lev2},\cite{Lipatov:1994xy,Faddeev:1994zg}. 
A similar open integrable spin chain has been found in kinematical regions of $n$-point maximally helicity violating (MHV) and planar ($N_c \to \infty$) amplitudes in MSYM~\cite{Lipatov:2009nt,Bartels:2011nz}.  
In this case Mandelstam cut contributions and the BFKL kernel in the adjoint representation play a 
fundamental role~\cite{Bartels:2008ce,Bartels:2008sc}.
Very recently, a novel relation between the BFKL equation in the forward limit and the sl(2) invariant XXX spin -1/2 chain~\cite{arXiv:1111.4553} has been unveiled.

The effective vertex in Eq.~\eqref{QCDlevvertex} is universal in the sense that it is independent of the nature of the external particles to which it couples. 
We have chosen two distinct quarks for simplicity, but it would be the same for only external gluons, for example. 
We follow a similar logic for the gravitational Einstein-Hilbert theory in the following section.

\section{Scalar-scalar scattering with graviton emission}

To minimize the number of contributing Feynman diagrams, in this section we analyze the gravitational scattering of two distinct scalars $\phi, \Phi$ 
with the emission of a graviton in the final state with polarization $\epsilon_{\mu\nu}(k)$:
\begin{eqnarray}
\phi(p)+\Phi(q)\longrightarrow \phi(p')+\Phi (q')+G(k,\epsilon).
\end{eqnarray}
We proceed with the calculation of the corresponding amplitude using 
algebraic codes~\cite{xperm} when expressions are lengthy. A novel exact effective vertex will be obtained whose MRK limit will 
coincide with the one calculated by Lipatov in his works on gravity. Our exact vertex for graviton emission depends on 
the particular choice of external particles while its MRK limit is universal. The tree-level amplitude involves
the computation of seven Feynman diagrams:
\begin{eqnarray}
\nonumber \\[0.2cm]
\mathcal{M}&\equiv&\hspace*{1cm}
\parbox{42mm}{
\begin{fmfgraph*}(90,60)
\fmfbottom{i1,d1,o1}
\fmftop{i2,d2,o2}
\fmfright{r1}
\fmflabel{$p$}{i2}
\fmflabel{$p'$}{o2}
\fmflabel{$(k,\epsilon)$}{r1}
\fmflabel{$q$}{i1}
\fmflabel{$q'$}{o1}
\fmf{plain}{i1,v1,v2,o1}
\fmf{plain}{i2,v3,v4,o2}
\fmf{phantom}{i1,o1}
\fmf{dbl_wiggly,tension=0}{v1,v3}
\fmf{dbl_wiggly,tension=0}{v4,r1}
\end{fmfgraph*} 
}\hspace*{0.5cm}+\hspace*{0.8cm}
\parbox{42mm}{
\begin{fmfgraph*}(90,60)
\fmfbottom{i1,d1,o1}
\fmftop{i2,d2,o2}
\fmfright{r1}
\fmflabel{$p$}{i2}
\fmflabel{$p'$}{o2}
\fmflabel{$(k,\epsilon)$}{r1}
\fmflabel{$q$}{i1}
\fmflabel{$q'$}{o1}
\fmf{plain}{i1,v1,v2,o1}
\fmf{phantom}{i1,o1}
\fmf{plain}{i2,v3,v4,o2}
\fmf{dbl_wiggly,tension=0}{v2,v4}
\fmf{dbl_wiggly,tension=0,rubout}{r1,v3}
\end{fmfgraph*} 
} \nonumber \\[0.9cm]
& & \hspace*{-0.4cm} +\hspace*{1cm}
\parbox{42mm}{
\begin{fmfgraph*}(90,60)
\fmfbottom{i1,d1,o1}
\fmftop{i2,d2,o2}
\fmfright{r1}
\fmflabel{$p$}{i2}
\fmflabel{$p'$}{o2}
\fmflabel{$(k,\epsilon)$}{r1}
\fmflabel{$q$}{i1}
\fmflabel{$q'$}{o1}
\fmf{plain}{i1,v2,o1}
\fmf{phantom}{i1,o1}
\fmf{plain}{i2,v4,o2}
\fmf{dbl_wiggly,tension=0}{v2,v4}
\fmf{dbl_wiggly,tension=0,rubout}{v4,r1}
\end{fmfgraph*} 
}\hspace*{0.5cm}+\hspace*{0.8cm}
\parbox{42mm}{
\begin{fmfgraph*}(90,60)
\fmfbottom{i1,d1,o1}
\fmftop{i2,d2,o2}
\fmfright{r1}
\fmflabel{$p$}{i2}
\fmflabel{$p'$}{o2}
\fmflabel{$(k,\epsilon)$}{r1}
\fmflabel{$q$}{i1}
\fmflabel{$q'$}{o1}
\fmf{plain}{i1,v1,v2,o1}
\fmf{phantom}{i1,o1}
\fmf{plain}{i2,v3,v4,o2}
\fmf{dbl_wiggly,tension=0}{v1,v3}
\fmf{dbl_wiggly,tension=0,rubout}{v2,r1}
\end{fmfgraph*} 
}
\label{eq:feyn_diag_grav}\nonumber\\[0.9cm]
& & \hspace*{-0.4cm} +\hspace*{1cm}
\parbox{42mm}{
\begin{fmfgraph*}(90,60)
\fmfbottom{i1,d1,o1}
\fmftop{i2,d2,o2}
\fmfright{r1}
\fmflabel{$p$}{i2}
\fmflabel{$p'$}{o2}
\fmflabel{$(k,\epsilon)$}{r1}
\fmflabel{$q$}{i1}
\fmflabel{$q'$}{o1}
\fmf{plain}{i1,v1,v2,o1}
\fmf{phantom}{i1,o1}
\fmf{plain}{i2,v3,v4,o2}
\fmf{dbl_wiggly,tension=0}{v2,v4}
\fmf{dbl_wiggly,tension=0,rubout}{v1,r1}
\end{fmfgraph*} 
}\hspace*{0.5cm}+\hspace*{0.8cm}
\parbox{42mm}{
\begin{fmfgraph*}(90,60)
\fmfbottom{i1,d1,o1}
\fmftop{i2,d2,o2}
\fmfright{r1}
\fmflabel{$p$}{i2}
\fmflabel{$p'$}{o2}
\fmflabel{$(k,\epsilon)$}{r1}
\fmflabel{$q$}{i1}
\fmflabel{$q'$}{o1}
\fmf{plain}{i1,v2,o1}
\fmf{phantom}{i1,o1}
\fmf{plain}{i2,v4,o2}
\fmf{dbl_wiggly,tension=0}{v2,v4}
\fmf{dbl_wiggly,tension=0,rubout}{v2,r1}
\end{fmfgraph*} 
}
\nonumber \\[0.9cm]
& & 
\hspace*{2.5cm}
+\hspace*{1cm}
\parbox{42mm}{
\begin{fmfgraph*}(90,60)
\fmfbottom{i1,d1,o1}
\fmftop{i2,d2,o2}
\fmfright{r1}
\fmflabel{$p$}{i2}
\fmflabel{$p'$}{o2}
\fmflabel{$q$}{i1}
\fmflabel{$q'$}{o1}
\fmf{plain}{i1,v1,o1}
\fmf{plain}{i2,v3,o2}
\fmf{dbl_wiggly,tension=0}{v1,v3}
\fmf{phantom}{i1,o1}
\end{fmfgraph*} 
}\hspace*{-2.6cm}
\parbox{42mm}{
\begin{fmfgraph*}(40,20)
\fmfleft{i1}
\fmfright{o1}
\fmflabel{$(k,\epsilon)$}{o1}
\fmf{dbl_wiggly,tension=1}{i1,o1}
\end{fmfgraph*} 
}
\hspace*{-2cm} 
\\ \nonumber
\end{eqnarray}
In the following we denote by $\mathcal{M}_{i}$ (with $i=1,\ldots,7$) the contribution of each of these 
diagrams which we have calculated using the Feynman rules listed in Appendix \ref{appendix_fr}.
In order to recast the long expressions in a more convenient way, we introduce the momenta $k_{1}$ and 
$k_{2}$ defined in Eq. \eqref{eq:new_momenta} and make use again of the Sudakov parametrization \eqref{eq:sudakov_paramet}.
Moreover, the graviton polarization tensor $\epsilon_{\mu\nu}(k)$ is written as
\begin{eqnarray}
\epsilon_{\mu\nu}(k)=\varepsilon_{\mu}(k)\varepsilon_{\nu}(k),
\end{eqnarray}
where $\varepsilon(k)\cdot\varepsilon(k)=0$ and $k\cdot\varepsilon(k)=0$. 
Using the last condition to write
\begin{eqnarray}
\varepsilon\cdot k_{1\perp}&=& \varepsilon\cdot k_{1\perp} -{1\over 2}\varepsilon\cdot k
\nonumber \\
&=&{1\over 2}\Big[
\varepsilon\cdot (k_{1\perp}+k_{2\perp})-(\alpha_{1}-\alpha_{2})\varepsilon\cdot p
-(\beta_{1}-\beta_{2})\varepsilon\cdot q
\Big],\\
\varepsilon\cdot k_{2\perp}&= & \varepsilon\cdot k_{2\perp} +{1\over 2}\varepsilon\cdot k
\nonumber \\
&=&{1\over 2}\Big[
\varepsilon\cdot(k_{1\perp}+k_{2\perp})+(\alpha_{1}-\alpha_{2})\varepsilon\cdot p
+(\beta_{1}-\beta_{2})\varepsilon\cdot q
\Big],
\end{eqnarray}
the total amplitude $\mathcal{M}$ 
can be shown to have the structure
\begin{eqnarray}
\mathcal{M}&=&[\varepsilon\cdot(k_{1\perp}+k_{2\perp})][\varepsilon\cdot(k_{1\perp}+k_{2\perp})]A_{kk}
+[\varepsilon\cdot(k_{1 \perp}+k_{2 \perp})](\varepsilon\cdot p)A_{kp} \nonumber \\
&+& [\varepsilon\cdot(k_{1 \perp}+k_{2 \perp})](\varepsilon\cdot q)A_{kq}+ (\varepsilon\cdot p)(\varepsilon\cdot p)
A_{pp}+ (\varepsilon\cdot q)(\varepsilon\cdot q)A_{qq} \nonumber\\
&+&  (\varepsilon\cdot p)(\varepsilon\cdot q)A_{pq}.
\label{eq:amplitude_general}
\end{eqnarray}
The six coefficients ${\cal A}_{ii}$ appearing in this expression 
are rational functions of the momenta whose explicit form is given in Appendix 
\ref{apB1}. The results are independent of the center of mass energy $s=(p+q)^{2}$ and can be written solely in terms of the Sudakov variables $\alpha_{1,2}$, $\beta_{1,2}$.

Before proceeding any further, we should point out that the computed amplitude is gauge invariant, {\it i.e.} 
using momentum conservation we have 
\begin{eqnarray}
k_{\mu}\mathcal{M}^{\mu\nu}=0=k_{\nu}\mathcal{M}^{\mu\nu},
\end{eqnarray}
where we have written $\mathcal{M}=\epsilon_{\mu\nu}\mathcal{M}^{\mu\nu}$.
As a further cross-check of our calculations, we note that the total amplitude also satisfies the Steinmannn 
relations \cite{Steinmann}. 
These are a consequence of unitarity and state that the amplitude cannot have simultaneous singularities, 
or multiple poles in energy variables, in overlapping channels. In our case the invariant masses associated with these two channels are
\begin{eqnarray}
s_{p'k}=(p'+k)^{2}, \hspace*{1cm} s_{q'k}=(q'+k)^{2},
\end{eqnarray}
which, in terms of the Sudakov variables, take the form
\begin{eqnarray}
s_{p'k}=-s(\alpha_{2}+\beta_{2}), \hspace*{1cm} s_{q'k}=s(\alpha_{1}+\beta_{1}).
\end{eqnarray}
In a preliminary version of the expressions in Appendix~\ref{apB1} it turned out that 
all six coefficients ${\cal A}_{ii}$ contained  
the potentially dangerous combination $(\alpha_{1}+\beta_{1})(\alpha_{2}+\beta_{2})$ in their  
denominators. We explicitly checked, however, that the numerators
vanish either when $\alpha_{1}\rightarrow -\beta_{1}$ or $\alpha_{2}\rightarrow-\beta_{2}$, cancelling out 
one of the factors in the denominator and leaving behind simple poles in $s_{p'k}$ or $s_{q'k}$. 
We have finally simplified our expressions to explicitly show the non-existence of these unphysical poles. 
This fulfillment of Steinmann relations provides a highly nontrivial test of our results.

As in the QCD case analyzed above, our aim is to decompose the total amplitude in terms of
gauge invariant combinations. 
Here we define
\begin{eqnarray}
\mathcal{M}_{\uparrow}=\mathcal{M}_{1}+\mathcal{M}_{2}+\mathcal{M}'
\hspace*{0.5cm} \mbox{and} \hspace*{0.5cm} 
\mathcal{M}_{\downarrow}=\mathcal{M}_{4}+\mathcal{M}_{5}+\mathcal{M}'',
\label{eq:mupdown}
\end{eqnarray}
where
\begin{eqnarray}
\mathcal{M}'={t\over t-t'}\Big(\mathcal{M}_{3}+\mathcal{M}_{7}\Big), \hspace*{1cm} 
\mathcal{M}''={t'\over t'-t}\Big(\mathcal{M}_{6}+\mathcal{M}_{7}\Big).
\end{eqnarray}
The total amplitude can be written as
\begin{eqnarray}
\mathcal{M}=\mathcal{M}_{\uparrow}+\mathcal{M}_{\downarrow}+\left({t'\over t'-t}\mathcal{M}_{3}+
{t\over t-t'}\mathcal{M}_{6}\right).
\end{eqnarray}
Remarkably, after a long calculation one can show that the last term cancels
\begin{eqnarray}
{t'\over t'-t}\mathcal{M}_{3}+
{t\over t-t'}\mathcal{M}_{6}=0.
\label{eq:funny_identity}
\end{eqnarray}
To qualitatively understand this result we notice that both the scalar-scalar-graviton and
scalar-scalar-graviton-graviton vertices given in Appendix~\ref{appendix_fr} are proportional to the
same kinematical factor $p_{\mu}q_{\nu}+p_{\nu}q_{\mu}$. After working out all the 
index contractions in $\mathcal{M}_{3}$ and $\mathcal{M}_{6}$ it is possible to find that 
the only difference in the contributions of these two diagrams lies in the denominator of the interchanged 
graviton. Diagramatically,
\begin{eqnarray}
\nonumber \\
t'\times \left(\hspace*{0.6cm}
\parbox{42mm}{
\begin{fmfgraph*}(90,60)
\fmfbottom{i1,d1,o1}
\fmftop{i2,d2,o2}
\fmfright{r1}
\fmflabel{$p$}{i2}
\fmflabel{$p'$}{o2}
\fmflabel{$(k,\epsilon)$}{r1}
\fmflabel{$q$}{i1}
\fmflabel{$q'$}{o1}
\fmf{plain}{i1,v2,o1}
\fmf{phantom}{i1,o1}
\fmf{plain}{i2,v4,o2}
\fmf{dbl_wiggly,tension=0}{v2,v4}
\fmf{dbl_wiggly,tension=0,rubout}{v4,r1}
\end{fmfgraph*} 
}
\right)=
t\times\left(\hspace*{0.6cm}
\parbox{42mm}{
\begin{fmfgraph*}(90,60)
\fmfbottom{i1,d1,o1}
\fmftop{i2,d2,o2}
\fmfright{r1}
\fmflabel{$p$}{i2}
\fmflabel{$p'$}{o2}
\fmflabel{$(k,\epsilon)$}{r1}
\fmflabel{$q$}{i1}
\fmflabel{$q'$}{o1}
\fmf{plain}{i1,v2,o1}
\fmf{phantom}{i1,o1}
\fmf{plain}{i2,v4,o2}
\fmf{dbl_wiggly,tension=0}{v2,v4}
\fmf{dbl_wiggly,tension=0,rubout}{v2,r1}
\end{fmfgraph*} 
}
\right),
\\ \nonumber
\end{eqnarray}
from where the cancellation in Eq.~\eqref{eq:funny_identity} follows.

We conclude that the amplitude can be written only in terms of the two contributions shown in 
Eq.~\eqref{eq:mupdown}
\begin{eqnarray}
\mathcal{M}=\mathcal{M}_{\uparrow}+\mathcal{M}_{\downarrow},
\end{eqnarray}
where each of the partial amplitudes $\mathcal{M}_{\uparrow}$, $\mathcal{M}_{\downarrow}$ have the 
structure shown in Eq.~\eqref{eq:amplitude_general}. Their corresponding coefficients are
given in Appendix~\ref{apB2}.
What makes this decomposition interesting is that both terms on the right-hand side are gauge
invariant independently, {\it i.e.} they satisfy the Ward identities
\begin{eqnarray}
k_{\mu}\mathcal{M}^{\mu\nu}_{\uparrow}=0, \hspace*{1cm}
k_{\mu}\mathcal{M}^{\mu\nu}_{\downarrow}=0,
\end{eqnarray}
where $\mathcal{M}_{\uparrow}\equiv\epsilon_{\mu\nu}\mathcal{M}^{\mu\nu}_{\uparrow}$ and 
$\mathcal{M}_{\downarrow}\equiv\epsilon_{\mu\nu}\mathcal{M}^{\mu\nu}_{\downarrow}$. As in the 
gauge theory case analyzed above, we can write the two gauge invariant contributions 
in terms of an effective vertex for the interaction of two on-shell scalars with one on-shell and one 
off-shell gravitons. The pictorial representation would be:
\begin{eqnarray}
\nonumber \\
\mathcal{M}_{\uparrow}=\hspace*{0.8cm}
\parbox{42mm}{
\begin{fmfgraph*}(90,60)
\fmfbottom{i1,d1,o1}
\fmftop{i2,d2,o2}
\fmfblob{.15w}{v4}
\fmfright{r1}
\fmflabel{$p$}{i2}
\fmflabel{$p'$}{o2}
\fmflabel{$(k,\epsilon)$}{r1}
\fmflabel{$q$}{i1}
\fmflabel{$q'$}{o1}
\fmf{plain}{i1,v2,o1}
\fmf{phantom}{i1,o1}
\fmf{plain}{i2,v4,o2}
\fmf{dbl_wiggly,tension=0}{v2,v4}
\fmf{dbl_wiggly,tension=0,rubout}{v4,r1}
\end{fmfgraph*} 
}
\hspace*{1.5cm} 
\mathcal{M}_{\downarrow}=\hspace*{0.8cm}
\parbox{42mm}{
\begin{fmfgraph*}(90,60)
\fmfbottom{i1,d1,o1}
\fmftop{i2,d2,o2}
\fmfright{r1}
\fmfblob{.15w}{v2}
\fmflabel{$p$}{i2}
\fmflabel{$p'$}{o2}
\fmflabel{$(k,\epsilon)$}{r1}
\fmflabel{$q$}{i1}
\fmflabel{$q'$}{o1}
\fmf{plain}{i1,v2,o1}
\fmf{phantom}{i1,o1}
\fmf{plain}{i2,v4,o2}
\fmf{dbl_wiggly,tension=0}{v2,v4}
\fmf{dbl_wiggly,tension=0,rubout}{v2,r1}
\end{fmfgraph*} 
}
\label{eq:twogravitytopologies}
\\[0.2cm]\nonumber
\end{eqnarray}

We find this result very interesting and are currently investigating the generalization of these effective 
diagrams to 
higher point amplitudes. If any iterative structure for higher order effective vertices could be found 
it would drastically help simplify loop calculations in Einstein-Hilbert gravity when using unitarity techniques. 

\subsection{Gravity as a double copy of QCD in multi-Regge kinematics}

We are now ready to address a very interesting point which has attracted quite a lot of attention 
in recent literature: in which sense are our effective vertices in gravity and QCD related? It is not 
possible for us to answer this question directly in general kinematics since, in order to simplify our 
calculations, we have operated with very particular external states, scalars in Einstein-Hilbert and 
quarks in QCD. This issue has been discussed by Bern and collaborators in a series of works mainly involving the mapping of amplitudes with only gluons in one side and only gravitons in the 
other~\cite{Bern:2002kj,Bern:2010ue,Bern:2010yg,Bern:2011ia}. 
We can, however, investigate our scattering amplitudes in multi-Regge kinematics and try to 
reproduce the results by Lipatov~\cite{Lipatov:1982vv,Lipatov:1982it,Lipatov:1991nf} where he found that the gluon and graviton emission vertices in MRK 
are indeed related as described in the following. 

Let us make use of the relations $k_1^2 = \beta_1 s = t$ and $k_2^2 = - \alpha_2 s = t'$ to rewrite the QCD MRK effective vertex of Eq.~\eqref{QCDlevvertex} in the form
\begin{eqnarray}
 {\Gamma^{\nu}_{\mu\sigma} } =  i g \eta_{\mu\sigma} \left[
 \left(\alpha_{1} - {2 \beta_1 \over   \beta_{2}}\right) p^{\nu}
+ \left(\beta_{2} + {2 \alpha_2 \over \alpha_{1}}\right) q^{\nu}
-(k_{1}+k_{2})_{\perp}^{\nu} \right] \equiv i g \eta_{\mu\sigma}  \Omega^\nu .
\end{eqnarray}
When looking for gravity as a simple double copy of the gauge theory amplitudes we find a difficulty in MRK. In this region, which corresponds to the limit of Sudakov 
parameters described in Eq.~\eqref{MRK1}, we have that $s_{p' k} 
= - \left(\alpha_2 + \beta_2 \right) s \simeq - \beta_2 s$ and $s_{q' k} = 
\left(\alpha_1 + \beta_1\right) s \simeq \alpha_1 s$, and can write
\begin{eqnarray}
 \Omega^\nu  \simeq  
 \left(\frac{s_{q' k} }{s} + {2 k_1^2 \over  s_{p' k}}\right) p^{\nu}
- \left(\frac{s_{p' k} }{s} + {2 k_2^2 \over  s_{q' k}}\right) q^{\nu}
-(k_{1}+k_{2})_{\perp}^{\nu}.
\end{eqnarray}
It is important to notice that, when written in this form, the naive double copy 
$\Omega^\mu \Omega^\nu$ has explicit unphysical double poles of the form $(s_{p' k} s_{q' k})^{-1}$ which correspond to simultaneous energy discontinuities in overlapping channels. The solution 
proposed by Lipatov was to introduce a subtraction term in order to fulfill Steinmann relations which reads
\begin{eqnarray}
{\cal N}^\mu &=& 2 \sqrt{k_1^2 k_2^2} \left(\frac{p^\mu}{s_{p' k} }-\frac{q^\mu}{s_{q' k}}\right) 
~ \simeq~ -2 i \sqrt{\beta_1 \alpha_2} \left(\frac{p^\mu}{\beta_2}+\frac{q^\mu}{\alpha_1}\right),
\end{eqnarray}
with the corresponding double copy being
\begin{eqnarray}
{\cal N}^\mu {\cal N}^\nu &\simeq& - 4 \beta_1 \alpha_2 \left(\frac{p^\mu p^\nu}{\beta_2^2}
+\frac{q^\mu q^\nu}{\alpha_1^2} + \frac{p^\mu q^\nu + q^\mu p^\nu}{\alpha_1 \beta_2}\right). 
\end{eqnarray}
What Lipatov showed is that the effective vertex for the emission of a graviton in Einstein-Hilbert gravity 
can be written in MRK as the following combination of QCD MRK effective vertices:
\begin{eqnarray}
\Omega^\mu \Omega^\nu - {\cal N}^\mu {\cal N}^\nu &=& 
(k_{1}+k_{2})_{\perp}^{\mu} (k_{1}+k_{2})_{\perp}^{\nu} 
+ \left[\left(\alpha_{1} -  {2 \beta_1 \over  \beta_{2}}\right)^2 + 4 \frac{\beta_1 \alpha_2}{\beta_2^2}\right] p^{\mu} p^\nu \nonumber\\
&&\hspace{-3cm}+ \left[ \left(\beta_{2}  + {2 \alpha_2 \over \alpha_{1}}\right)^2 + 4 \frac{\beta_1 \alpha_2}{\alpha_1^2}\right]q^{\mu} q^{\nu}
+ \left[ \left(\alpha_{1} -{2 \beta_1 \over  \beta_{2}}\right)
\left(\beta_{2}  + {2 \alpha_2 \over \alpha_{1}}\right) + 4 \frac{\beta_1 \alpha_2}{\alpha_1 \beta_2} \right] \left(p^\mu q^\nu + q^\mu p^\nu\right) \nonumber\\
&&\hspace{-3cm}- \left(\alpha_{1} - {2 \beta_1 \over  \beta_{2}}\right)
 \left(p^\mu k^\nu + k^\mu p^\nu\right) 
 - \left(\beta_{2}  + {2 \alpha_2 \over \alpha_{1}}\right)
 \left(q^\mu k^\nu + k^\mu q^\nu\right). 
\label{GGNN} 
\end{eqnarray}

Let us see if this complicated structure is present in our exact calculations. For this purpose it is 
needed to find the corresponding coefficients in the expansion of our 
Eq.~\eqref{eq:amplitude_general} which now we write in the form
\begin{eqnarray}
{\cal M}^{\mu \nu} &=& (k_{1}+k_{2})_{\perp}^{\mu} (k_{1}+k_{2})_{\perp}^{\nu} 
+ {\cal A}_{pp} \, p^{\mu} p^\nu +{\cal A}_{qq} \, q^{\mu} q^{\nu} 
+ {\cal A}_{pq} \left(p^\mu q^\nu + q^\mu p^\nu\right) \nonumber\\
&+&{\cal A}_{kp}  \left(p^\mu k^\nu + k^\mu p^\nu\right) 
 + {\cal A}_{kq} \left(q^\mu k^\nu + k^\mu q^\nu\right), 
\label{rearranged} 
\end{eqnarray}
where\footnote{The process dependent impact factors can be evaluated from the MRK limit of the normalization factor $A_{kk}$.} 
${\cal A}_{ii} = A_{ii} / A_{kk}$.

Using the results for the exact gravitational amplitude given in the Appendix and operating in the 
MRK of Eq.~\eqref{MRK1}, the Taylor series expansion of each of the coefficients in 
Eq.~\eqref{rearranged} is as follows:

\begin{eqnarray}
{\cal A}_{pp} &=& \alpha _1^2+{\cal O}\left(\alpha _1^3\right)+\beta _1
   \left[-\frac{4 \alpha _1}{\beta _2}+\frac{4 \alpha _1^2}{\beta
   _2}+{\cal O}\left(\alpha _1^3\right)\right]+\beta _1^2 \left[\frac{4}{\beta
   _2^2}-\frac{12 \alpha _1}{\beta _2^2}+\frac{8 \alpha _1^2}{\beta
   _2^2}+{\cal O}\left(\alpha _1^3\right)\right] \nonumber\\
   &+&{\cal O}\left(\beta_1^3\right) + \alpha _2 \Bigg\{-2 \alpha _1+{\cal O}\left(\alpha
   _1^3 \right)+\left[\frac{8 \beta _2+4}{\beta _2^2}-\frac{\left(8
   \beta _2+4\right) \alpha _1}{\beta _2^2}+{\cal O}\left(\alpha _1^3\right)\right]
   \beta _1  \nonumber\\
   &+& \left[\frac{4 \left(\beta _2+1\right)}{\beta _2^2 \alpha
   _1}+\frac{4-8 \beta _2^2}{\beta _2^3}+\frac{4 \alpha _1}{\beta _2}-\frac{4
   \left(\beta _2+1\right) \alpha _1^2}{\beta _2^3}+{\cal O}\left(\alpha
   _1^3\right)\right] \beta _1^2+{\cal O}\left(\beta
   _1^3\right)\Bigg\} +{\cal O}\left(\alpha _2^2\right) \nonumber\\
   &\simeq& \alpha _1^2-\frac{4 \alpha _1 \beta _1}{\beta _2}+\frac{4 \beta _1^2}{\beta _2^2}
   +\frac{4 \alpha _2 \beta _1}{\beta_2^2} + \dots 
\end{eqnarray}
\begin{eqnarray}
{\cal A}_{qq} &=& \beta _2^2+{\cal O}\left(\beta _2^3\right) +\beta _1 \left[-2
   \beta _2+{\cal O}\left(\beta _2^3\right)\right] +{\cal O}\left(\beta
   _1^2\right) \nonumber\\
   &+&\alpha _2 \Bigg\{ \frac{4 \beta _2}{\alpha
   _1}+\frac{4 \beta _2^2}{\alpha _1}+{\cal O}\left(\beta
   _2^3\right) +\left[\frac{4-8 \alpha _1}{\alpha _1^2}+\frac{\left(4-8
   \alpha _1\right) \beta _2}{\alpha _1^2}+{\cal O}\left(\beta _2^3\right)\right]
   \beta _1+{\cal O}\left(\beta _1^2\right)\Bigg\} \nonumber\\
   &+&\alpha _2^2
   \Bigg\{ \frac{4}{\alpha _1^2}+\frac{12 \beta _2}{\alpha _1^2}+\frac{8
   \beta _2^2}{\alpha _1^2}+{\cal O}\left(\beta _2^3\right)  \nonumber\\
   &+&  \left[\frac{4-4
   \alpha _1}{\alpha _1^2 \beta _2}+\frac{4-8 \alpha _1^2}{\alpha
   _1^3}-\frac{4 \beta _2}{\alpha _1}+\frac{4 \left(\alpha _1-1\right) \beta
   _2^2}{\alpha _1^3}+{\cal O}\left(\beta _2^3\right)\right] \beta _1+{\cal O}\left(\beta
   _1^2\right)\Bigg\}+{\cal O}\left(\alpha _2^3\right)\nonumber\\
   &\simeq& \beta _2^2 +\frac{4 \alpha _2 \beta _2}{\alpha
   _1}+ \frac{4 \alpha _2 \beta _1}{\alpha _1^2}+\frac{4 \alpha _2^2}{\alpha _1^2}+ \dots 
\end{eqnarray}
\begin{eqnarray}
{\cal A}_{pq} &=& \alpha _1 \beta _2+{\cal O}\left(\beta _2^2\right) +\left(\alpha
   _1-2\right) \beta _1+{\cal O}\left(\beta _1^2\right) + \alpha _2
   \left[2+\beta _2+{\cal O}\left(\beta _2^2\right) +\beta
   _1+{\cal O}\left(\beta _1^2\right)\right] +{\cal O}\left(\alpha _2^2\right)\nonumber\\
   &\simeq&  \alpha _1 \beta _2-2 \beta _1+2 \alpha _2+ \dots \\
{\cal A}_{kp} &=& -\alpha _1+\beta _1 \left[\frac{2-2 \alpha _1}{\beta _2}+{\cal O}\left(\beta
   _2^2\right) +{\cal O}\left(\beta _1^2\right)\right]+{\cal O}\left(\alpha
   _2^1\right)\nonumber\\
   &\simeq& -\alpha _1 + \frac{2 \beta _1}{\beta _2} + \dots \\
{\cal A}_{kq} &=& -\beta _2+{\cal O}\left(\beta _2^2\right) +{\cal O}\left(\beta
   _1^1\right) +\alpha _2 \left[ -\frac{2}{\alpha _1}-\frac{2
   \beta _2}{\alpha _1}+{\cal O}\left(\beta _2^2\right) +{\cal O}\left(\beta
   _1^1\right) \right] + {\cal O}\left(\alpha _2^2\right)\nonumber\\
   &\simeq& -\beta _2 -\frac{2 \alpha _2}{\alpha _1}+ \dots
\end{eqnarray}
These coefficients calculated in MRK are therefore in exact agreement with those of 
Eq.~\eqref{GGNN}. This is a highly non-trivial check of our calculation which sheds light on the deep 
relation between Einstein-Hilbert gravity and gauge theories. 

The procedure we have followed in this work to cross-check our calculations can be applied to amplitudes 
with an arbitrary number of loops and external legs. We have found that the 
representation of the exact amplitudes in Sudakov variables facilitates the application of the MRK limit 
and the comparison with the iterated form of amplitudes valid in this region. This adds to the more 
standard checks related to gauge invariance and agreement with the Steinmann relations.

\section{Conclusions}

In this work the tree level five-point amplitude for the scattering of two distinct scalars 
with a graviton in the final state has been evaluated considering Einstein-Hilbert gravity as a usual gauge theory. 
The calculation has been performed in exact kinematics with the final result expressed in Sudakov variables. 
We have tested not only the gauge invariance of the full amplitude but also the lack of simultaneous singularities 
in overlapping channels, in agreement with Steinmann relations. 
 
We have found the remarkable result that, due to the subtle cancellation shown in Eq.~(\ref{eq:funny_identity}), 
the full amplitude can be written as the sum of only the two gauge invariant topologies given in  
Eq.~(\ref{eq:twogravitytopologies}), both written in terms of the same effective vertex for graviton emission off 
a scalar line together with an off-shell graviton connecting with the other distinct scalar line. 
A natural expansion for a general five-point amplitude in gravity is given in Eq.~(\ref{eq:amplitude_general}) 
with the coefficients for our full result being shown in Appendix~\ref{apB1}. The coefficients for the 
separation into our two novel effective topologies are explicitly written in Appendix~\ref{apB2}. 
This new structure in terms of effective vertices will be useful to simplify and streamline the calculation of 
higher order corrections in gravity, reducing in a great amount the number of possible topologies contributing 
to a given process. It is likely that new effective topologies will appear as the number of 
external legs in the amplitude increases. The possible relations of the new effective vertices with the 
lower order ones  will be the subject of our future investigations. 

In order to investigate the interesting link between gravity and gauge theories we have also offered a detailed derivation of the QCD amplitude with four external quarks and one gluon. In this case a similar 
separation into effective topologies as in Einstein-Hilbert gravity appears. We have reproduced the results 
obtained by Lipatov many years ago by showing that the graviton emission vertex in multi-Regge kinematics 
can be written as the product of two gluon emission vertices in QCD also in the same limit, with an additional 
subtraction needed to fulfill the Steinmann relations. It will be interesting to generalize this calculation to 
supersymmetric theories, also investigated in Lipatov's works, and the interpretation of these 
results in terms of string theory. The connection between the results here presented and the 
puzzling duality between color and kinematics recently proposed 
in~\cite{Bern:2010ue,Bern:2010yg,Bern:2011ia} is the subject of current investigations.

\section*{Acknowledgments} 

We thank Luis {\' A}lvarez-Gaum{\' e}, Zvi Bern and  Lev Lipatov for useful discussions. 
ASV acknowledges partial support from the European Comission under contract LHCPhenoNet (PITN-GA-2010-264564), 
the Comunidad de Madrid through Proyecto HEPHACOS ESP-1473, and MICINN (FPA2010-17747). The work of ESC 
has been supported by a Spanish Government FPI Predoctoral Fellowship and grant FIS2009-07238. 
MAVM acknowledges partial support from Spanish Government grants FPA2009-10612 and FIS2009-07238, 
Basque Government Grant IT-357-07 and Spanish Consolider-Ingenio 2010 Programme CPAN (CSD2007-00042).
ESC and MAVM thank the Instituto de F\'{\i}sica Te\'orica UAM/CSIC for hospitality during the completion of
this work. Finally, we thank the CERN Theory Unit where the final stages of this investigation took place.


\appendix

\section{Feynman rules}
\label{appendix_fr}

In this Appendix we list the Feynman rules used in the calculation of the gravitational amplitude
\eqref{eq:feyn_diag_qcd}.  Our action, where there are two types of distinct massless scalars denoted by $\phi$ and $\Phi$, reads
\begin{equation}
	S= \int d^4x\sqrt{-g}\left(-\frac{1}{\kappa^2 }R+{1\over 2}
	g^{\mu \nu }\partial _{\mu }\phi \partial _{\nu }\phi 
	+{1\over 2} g^{\mu \nu }\partial _{\mu }\Phi \partial _{\nu }\Phi  \right),
	\label{eq:fulllagrangian}
\end{equation}
with $\kappa^{2}=16\pi G_{N}$.
To simplify the calculations we follow~\cite{LandauLifschitzClassicalFields} and make use of the identity
\begin{equation}
	\sqrt{|g|}R [g]=\sqrt{|g|}g^{\mu \nu }\left(\Gamma _{\mu \beta }^{\alpha }\Gamma _{\nu \alpha }^{\beta }-\Gamma _{\mu \nu }^{\alpha }\Gamma _{\alpha \beta }^{\beta }\right) + \text{total derivatives.} 
\end{equation}
Writing $g_{\mu\nu}=\eta_{\mu\nu}+\kappa h_{\mu\nu}$ and using the de Donder gauge, 
\begin{eqnarray}
\partial_{\mu}h^{\mu}_{\nu}={1\over 2}\partial_{\nu}h^{\alpha}_{\alpha}, 
\label{eq:dedonder}
\end{eqnarray}
the propagators are given by
\begin{eqnarray}
\parbox{42mm}{
\begin{fmfgraph*}(90,60)
\fmfleft{l1}
\fmfright{r1}
\fmf{plain,label=$p\rightarrow$}{l1,r1}
\end{fmfgraph*} 
} & \Longrightarrow & {i\over p^{2}+i\epsilon}, 
\\
\parbox{42mm}{
\begin{fmfgraph*}(90,60)
\fmfleft{l1}
\fmfright{r1}
\fmflabel{$\alpha\beta$}{l1}
\fmflabel{$\sigma\lambda$}{r1}
\fmf{dbl_wiggly,label=$p\rightarrow$}{l1,r1}
\end{fmfgraph*} 
}& \Longrightarrow & {i\over p^{2}+i\epsilon}\Big(\eta_{\alpha\sigma}\eta_{\beta\lambda}
+\eta_{\alpha\lambda}\eta_{\beta\sigma}-\eta_{\alpha\beta}\eta_{\sigma\lambda}\Big). 
\end{eqnarray}
At ${\cal O} (\kappa^{2})$ there exist two interaction vertices between scalars and gravitons. The first one is a 
scalar-scalar-graviton vertex of the form\\
\begin{eqnarray}
\parbox{42mm}{
\begin{fmfgraph*}(90,60)
\fmfleft{l1,l2}
\fmfright{r1}
\fmf{plain,label=$\!\!\!\!\nearrow p$}{l1,v}
\fmf{plain,label=$\!\!\!\!\searrow q$}{v,l2}
\fmflabel{$\alpha\beta$}{r1}
\fmf{dbl_wiggly,label=$\leftarrow k$}{v,r1}
\end{fmfgraph*} 
} & \Longrightarrow & -{i\kappa\over 2}(p_{\mu}q_{\nu}+p_{\nu}q_{\mu})\left(
-\eta^{\mu\alpha}\eta^{\nu\beta}+{1\over 2}\eta^{\mu\nu}\eta^{\alpha\beta}\right).
\label{eq:ssg_fr}
\end{eqnarray}\\
The second one is a scalar-scalar-graviton-graviton vertex:\\
\begin{eqnarray}
\parbox{42mm}{
\begin{fmfgraph*}(90,60)
\fmfleft{l1,l2}
\fmfright{r1,r2}
\fmf{plain,label=$\!\!\!\!\nearrow p$}{l1,v}
\fmf{plain,label=$\!\!\!\!\searrow q$}{v,l2}
\fmflabel{$\alpha\beta$}{r1}
\fmflabel{$\sigma\lambda$}{r2}
\fmf{dbl_wiggly}{v,r1}
\fmf{dbl_wiggly}{r2,v}
\end{fmfgraph*} 
} & \Longrightarrow & -{i\kappa^{2}\over 4}(p_{\mu}q_{\nu}+p_{\nu}q_{\mu})
\left(\mathcal{I}^{\alpha\beta}_{\,\,\,\,\,\,\,\rho\zeta}
\mathcal{I}^{\sigma\lambda}_{\,\,\,\,\,\,\,\,\delta\gamma}+
\mathcal{I}^{\alpha\beta}_{\,\,\,\,\,\,\,\,\delta\gamma}
\mathcal{I}^{\sigma\lambda}_{\,\,\,\,\,\,\,\,\rho\zeta}
\right)  
\label{eq:ssgg_fr} 
\\
 &  & \,\,\,\,\times\left[
2 \eta^{\gamma\nu} \eta^{\zeta\delta} \eta^{\mu\rho} -  \eta^{\gamma\nu} \eta^{\mu\delta} \eta^{\rho\zeta} + \eta^{\mu\nu} \left(-{1\over 2} \eta^{\zeta\gamma} \eta^{\rho\delta} + {1\over 4} \eta^{\delta\gamma} \eta^{\rho\zeta}\right)
\right], 
\nonumber
\end{eqnarray}
where we have introduced the symmetrizer
\begin{eqnarray}
\mathcal{I}_{\alpha\beta;\gamma\delta}={1\over 2}\Big(\eta_{\alpha\gamma}
\eta_{\beta\delta}+\eta_{\alpha\delta}\eta_{\beta\gamma}\Big). 
\end{eqnarray}
Finally, there is the three-graviton vertex
\begin{eqnarray}
\nonumber \\
\parbox{42mm}{
\begin{fmfgraph*}(90,60)
\fmfleft{l1,l2}
\fmfright{r1}
\fmf{dbl_wiggly,label=$\!\!\!\!\nearrow p$}{l1,v}
\fmf{dbl_wiggly,label=$\!\!\!\!\searrow q$}{v,l2}
\fmflabel{$\alpha\beta$}{r1}
\fmflabel{$\mu\nu$}{l2}
\fmflabel{$\gamma\delta$}{l1}
\fmf{dbl_wiggly,label=$\leftarrow k$}{v,r1}
\end{fmfgraph*} 
} & \Longrightarrow & i\kappa  \Big(p_{\sigma}q_{\lambda}
\mathcal{I}^{\mu\nu}_{\,\,\,\,\,\,\,\,\zeta\xi}
\mathcal{I}^{\gamma\delta}_{\,\,\,\,\,\,\,\,\rho\chi}
\mathcal{I}^{\alpha\beta}_{\,\,\,\,\,\,\,\,\tau\upsilon}
+p_{\lambda}q_{\sigma}
\mathcal{I}^{\gamma\delta}_{\,\,\,\,\,\,\,\,\zeta\xi}
\mathcal{I}^{\mu\nu}_{\,\,\,\,\,\,\,\,\rho\chi}
\mathcal{I}^{\alpha\beta}_{\,\,\,\,\,\,\,\,\tau\upsilon}
\nonumber\\[-0.4cm]
 & & \,\,
 +\,\,k_{\sigma}q_{\lambda}
\mathcal{I}^{\mu\nu}_{\,\,\,\,\,\,\,\,\zeta\xi}
\mathcal{I}^{\alpha\beta}_{\,\,\,\,\,\,\,\,\rho\chi}
\mathcal{I}^{\gamma\delta}_{\,\,\,\,\,\,\,\,\tau\upsilon}
+
k_{\lambda}q_{\sigma}
\mathcal{I}^{\alpha\beta}_{\,\,\,\,\,\,\,\,\zeta\xi}
\mathcal{I}^{\mu\nu}_{\,\,\,\,\,\,\,\,\rho\chi}
\mathcal{I}^{\gamma\delta}_{\,\,\,\,\,\,\,\,\tau\upsilon} 
\nonumber\\[0.2cm]
 & & \,\,
 +\,\,k_{\sigma}p_{\lambda}
\mathcal{I}^{\gamma\delta}_{\,\,\,\,\,\,\,\,\zeta\xi}
\mathcal{I}^{\alpha\beta}_{\,\,\,\,\,\,\,\,\rho\chi}
\mathcal{I}^{\mu\nu}_{\,\,\,\,\,\,\,\,\tau\upsilon}
+
k_{\lambda}p_{\sigma}
\mathcal{I}^{\alpha\beta}_{\,\,\,\,\,\,\,\,\zeta\xi}
\mathcal{I}^{\gamma\delta}_{\,\,\,\,\,\,\,\,\rho\chi}
\mathcal{I}^{\mu\nu}_{\,\,\,\,\,\,\,\,\tau\upsilon} \Big)
\nonumber \\[0.2cm]
& & \,\,\times\,\, \mathscr{G}^{\lambda;\zeta\xi;\sigma;\rho\chi;\tau\upsilon},
\end{eqnarray}
where the last factor is given by
\begin{eqnarray}
\hspace*{-0.5cm} 
\mathscr{G}^{\lambda;\zeta\xi;\sigma;\rho\chi;\tau\upsilon}&=&-\frac{1}{2} \eta^{\zeta\tau} \eta^{\lambda\chi} \eta^{\xi\sigma} \eta^{\rho\nu} - \frac{1}{4} \eta^{\zeta\tau} \eta^{\lambda\sigma} \eta^{\xi\nu} \eta^{\rho\chi} + \frac{1}{4} \eta^{\zeta\sigma} \eta^{\lambda\chi} \eta^{\xi\rho} \eta^{\tau\nu} - \frac{1}{8} \eta^{\zeta\rho} \eta^{\lambda\sigma} \eta^{\xi\chi} \eta^{\tau\nu} \nonumber \\[0.1cm]
&-&  \eta^{\zeta\chi} \eta^{\lambda\rho} \eta^{\xi\nu} \eta^{\tau\sigma} + \frac{1}{2} \eta^{\zeta\tau} \eta^{\lambda\sigma} \eta^{\xi\rho} \eta^{\chi\nu} + \frac{1}{4} \eta^{\zeta\rho} \eta^{\lambda\nu} \eta^{\tau\sigma} \eta^{\chi\xi} + \frac{1}{2} \eta^{\zeta\xi} \eta^{\lambda\rho} \eta^{\sigma\nu} \eta^{\chi\tau}.
\end{eqnarray}
When the vertex is fully expanded it contains $3!\times 2^3 \times 8= 384$ terms. To simplify its form, 
momentum conservation together with the de Donder condition have been used. As far as we know, this is the simplest 
form for this vertex in the literature (compare, {\it e.g.}, with~\cite{DeWitt:1967uc}).

\section{Explicit form of the gravitational amplitude}
\label{appendixB}

\subsection{The full five-point amplitude}
\label{apB1}

To avoid cluttering the text with long expressions, here we
have collected the full form of the coefficients of the amplitude shown in Eq. \eqref{eq:amplitude_general}. 
They are expressed in terms of the Sudakov decomposition for the momenta $k_{1}$, $k_{2}$ given in Eq. 
\eqref{eq:sudakov_paramet}
\begin{eqnarray}
A_{kk} &=& \frac{i \kappa ^3 }{8} \Bigg\{\frac{1}{\beta_1 \alpha_2} 
- \frac{\left(1+\beta_1\right)}{\beta_1 \left(\alpha_1+\beta_1\right)}
- \frac{\left(1-\alpha_2\right)}{\alpha_2 \left(\alpha_2+\beta_2\right)}
\Bigg\},
\\
A_{kp}&=& \frac{i \kappa ^3 }{8 } \Bigg\{-\frac{\left(\alpha _1-\alpha
   _2\right){}^2}{\alpha _1 \alpha _2 \beta
   _1}-\frac{\left(\alpha _1-1\right)
   \left(\alpha _1+\alpha _2\right)}{\alpha _1
   \left(\alpha _1+\beta
   _1\right)}+\frac{\left(\alpha _2-1\right)
   \left(\alpha _1+\alpha _2-2\right)}{\alpha
   _2 \left(\alpha _2+\beta _2\right)} \Bigg\},
   \\
A_{kq}&=&  \frac{i \kappa ^3 }{8 }  \Bigg\{-\frac{\left(\beta _1-\beta
   _2\right){}^2}{\alpha _2 \beta _1 \beta
   _2}-\frac{\left(\beta _1+1\right)
   \left(\beta _1+\beta _2+2\right)}{\beta _1
   \left(\alpha _1+\beta _1\right)}+\frac{\left(\beta _2+1\right)
   \left(\beta _1+\beta _2\right)}{\beta _2
   \left(\alpha _2+\beta
   _2\right)}\Bigg\}, 
\\
A_{pp}&=&  \frac{i \kappa ^3 }{8 } \Bigg\{ \frac{\left(\alpha _1-\alpha
   _2\right)^3}{\alpha _1 \alpha _2 \beta
   _1}+\frac{4 \left(\alpha
   _1-1\right) \left(\alpha _1-\alpha
   _2-1\right)}{\alpha _2 \left(\beta _2-\beta
   _1\right)}\nonumber\\
   &-&\frac{\left(\alpha _1-1\right)
   \left(\alpha _1+\alpha
   _2\right)^2}{\alpha _1 \left(\alpha
   _1+\beta _1\right)}+\frac{\left(\alpha
   _2-1\right) \left(\alpha _1+\alpha
   _2-2\right)^2}{\alpha _2 \left(\alpha
   _2+\beta _2\right)}\Bigg\},
   \\
A_{qq}&=&  \frac{i \kappa ^3 }{8 } \Bigg\{-\frac{\left(\beta _1-\beta
   _2\right)^3}{\alpha _2 \beta _1 \beta
   _2}-\frac{4
   \left(\beta _1-\beta _2-1\right)
   \left(\beta _2+1\right)}{\left(\alpha
   _1-\alpha _2\right) \beta _1} \nonumber\\
   &-&\frac{\left(\beta _1+1\right)
   \left(\beta _1+\beta _2+2\right)^2}{\beta
   _1 \left(\alpha _1+\beta _1\right)}+\frac{\left(\beta _2+1\right)
   \left(\beta _1+\beta _2\right)^2}{\beta
   _2 \left(\alpha _2+\beta
   _2\right)}\Bigg\},
   \\
A_{pq} &=& \frac{i \kappa ^3 }{8} 
   \Bigg[\left(\alpha _1+\alpha _2-2\right) \left(\beta _1-\alpha
   _2\right)+\left(\alpha _1+\alpha _2\right) \left(\alpha _2+\beta
   _2\right)\Bigg] \nonumber\\
   &\times& \Bigg\{\frac{\alpha _1-\alpha _2}{\alpha _1 \alpha _2 \beta_1}+\frac{1-\alpha _1}{\alpha _1 \left(\alpha_1+\beta _1\right)}
   +\frac{\alpha _2-1}{\alpha _2 \left(\alpha _2+\beta _2\right)}\Bigg\}. 
\end{eqnarray}

\subsection{The two topologies contributing to the full amplitude}
\label{apB2}

Next we list the coefficients corresponding to the partial amplitude $\mathcal{M}_{\uparrow}$
\begin{eqnarray}
A_{kk}^{\uparrow}  &=& \frac{i \kappa ^3 }{8} \Bigg\{\frac{1}{\alpha _2
   \beta _1}-\frac{\left(\beta _1+1\right)}{\beta _1
   \left(\alpha _2+\beta _1\right)}+\frac{\alpha _2-1}{\alpha _2 \left(\alpha
   _2+\beta _2\right)}\Bigg\}, 
   \\
A_{kp}^{\uparrow} &=& \frac{i \kappa ^3 }{8 \alpha _2} \Bigg\{\frac{\left(\alpha _2-1\right) \left(\alpha _1+\alpha
   _2-2\right)}{\alpha _2+\beta _2}-\frac{\alpha _1+\alpha _2 \left(2 \alpha
   _2-3\right)}{\alpha _2+\beta _1}\Bigg\}, 
   \\
A_{kq}^{\uparrow}&=& {i\kappa^{3}\over 8\alpha_{2}}\left\{{1\over \alpha_{2}+\beta_{1}}+{(1-\alpha_{2})(\alpha_{2}-\beta_{1})\over (\alpha_{2}+\beta_{1})(\alpha_{2}+\beta_{2})}\right\} ,
\\
A_{pp}^{\uparrow}&=& \frac{i \kappa^{3}}{8 \alpha _2} \Bigg\{\frac{\left(\alpha _2-1\right) \left(\alpha _1+\alpha
   _2-2\right){}^2}{\alpha _2+\beta _2}+\frac{\left(\alpha _1-3 \alpha
   _2\right) \Big[\alpha _2^2+\left(\alpha _1-3\right) \alpha _2+\alpha
   _1\Big]}{\alpha _2+\beta _1} \nonumber\\
   &-&\frac{4 \left(\alpha _1-1\right)
   \left(1-\alpha _1+\alpha _2\right)}{\beta _2-\beta _1} \Bigg\},
   \\
A_{qq}^{\uparrow} &=& {i\kappa^{3}\over 8\alpha_{2}}(\beta_{1}-\beta_{2})(\beta_{1}-\beta_{2}-2\alpha_{2})\left\{
{\alpha_{2}+1\over \alpha_{2}+\beta_{1}}+{(1-\alpha_{2})(\alpha_{2}-\beta_{1})\over (\alpha_{2}+\beta_{1})(\alpha_{2}+\beta_{2})}\right\}, 
\\
%
%
A_{pq}^{\uparrow} &=& {i\kappa^{3}\over 8}\Bigg\{{(\alpha_{2}-1)\Big[\beta_{1}(\alpha_{1}+\alpha_{2}-2)-2\beta_{2}(\alpha_{2}-1)
	-\alpha_{2}(\alpha_{1}-4)-3\alpha_{2}^{2}\Big]\over \alpha_{2}(\alpha_{2}+\beta_{2})} \nonumber \\
	&+&{\alpha_{1}\Big[\alpha_{2}^{2}+(\beta_{2}+3)\alpha_{2}+\beta_{2}\Big]+\alpha_{2}\Big[
	-3\beta_{2}+\alpha_{2}(3\alpha_{2}-\beta_{2}-9)
	\Big]
	\over \alpha_{2}(\alpha_{2}+\beta_{1})}
	\Bigg\}
\end{eqnarray}
and those of $\mathcal{M}_{\downarrow}$
\begin{eqnarray}
A_{kk}^{\downarrow}  &=& \frac{i \kappa ^3 }{8} \Bigg\{-\frac{1}{\alpha _1
   \beta _1}+\frac{1-\alpha _1}{\alpha _1 \left(\alpha
   _1+\beta _1\right)}+\frac{1+\beta _1}{\beta _1
   \left(\alpha _2+\beta _1\right)}\Bigg\},
   \\
A_{kp}^{\downarrow}  &=& \frac{i \kappa ^3 }{8} \Bigg\{-\frac{\left(\alpha _1-\alpha
   _2\right){}^2}{\alpha _1 \alpha _2 \beta
   _1}-\frac{\left(\alpha _1-1\right)
   \left(\alpha _1+\alpha _2\right)}{\alpha _1
   \left(\alpha _1+\beta
   _1\right)}+\frac{\alpha _1+\alpha _2
   \left(2 \alpha _2-3\right)}{\alpha _2
   \left(\alpha _2+\beta _1\right)}\Bigg\},
   \\
A_{kq}^{\downarrow}  &=& \frac{i \kappa ^3 }{8} \Bigg\{\frac{\beta _1 \left(2 \beta _1+3\right)-\beta
   _2}{\beta _1 \left(\alpha _2+\beta
   _1\right)}-\frac{\left(\beta _1+1\right)
   \left(\beta _1+\beta _2+2\right)}{\beta _1
   \left(\alpha _1+\beta _1\right)}\Bigg\},
   \\
A_{pp}^{\downarrow}  &=& \frac{i \kappa ^3 }{8} \Bigg\{\frac{\left(\alpha _1-\alpha
   _2\right){}^3}{\alpha _1 \alpha _2 \beta
   _1}-\frac{\left(\alpha _1-1\right)
   \left(\alpha _1+\alpha
   _2\right){}^2}{\alpha _1 \left(\alpha
   _1+\beta _1\right)} \nonumber \\
   &-&\frac{\left(\alpha _1-3
   \alpha _2\right) \Big[\alpha
   _2^2+\left(\alpha _1-3\right) \alpha
   _2+\alpha _1\Big]}{\alpha _2 \left(\alpha
   _2+\beta _1\right)}\Bigg\},
   \\
A_{qq}^{\downarrow}  &=& \frac{i \kappa ^3 }{8} \Bigg\{-\frac{\left(\beta _1+1\right) \left(\beta
   _1+\beta _2+2\right){}^2}{\beta _1
   \left(\alpha _1+\beta _1\right)}-\frac{4
   \left(\beta _1-\beta _2-1\right)
   \left(\beta _2+1\right)}{\left(\alpha
   _1-\alpha _2\right) \beta _1} 
   \nonumber
   \\
   &+&\frac{\left(3
   \beta _1-\beta _2\right) \Big[\beta _1
   \left(\beta _1+\beta _2+3\right)-\beta
   _2\Big]}{\beta _1 \left(\alpha _2+\beta
   _1\right)}\Bigg\},
\\
A_{pq}^{\downarrow}  &=& \frac{i \kappa ^3 }{8} \Bigg\{\frac{\left(\beta _1+1\right) \Big[\beta _1
   \left(-\alpha _2+\beta _2+4\right)+2 \alpha
   _1 \left(\beta _1+1\right)-\alpha _2
   \left(\beta _2+2\right)+3 \beta
   _1^2\Big]}{\beta _1 \left(\alpha _1+\beta
   _1\right)} \nonumber\\
   &+&\frac{\alpha _1 \Big[\beta
   _1^2-\left(\beta _2+3\right) \beta _1+\beta
   _2\Big]+\beta _1 \Big[3 \beta _2-\beta
   _1 \left(3 \beta _1+\beta
   _2+9\right)\Big]}{\beta _1 \left(\alpha
   _2+\beta _1\right)}\Bigg\}.
\end{eqnarray}

\end{fmffile}

\end{document}